# Conformity in virtual environments: a hybrid neurophysiological and psychosocial approach

Serena Coppolino Perfumi[1], Chiara Cardelli[1], Franco Bagnoli[3 4,] Andrea Guazzini[2 3]

[1] Neurofarba Department (Neuroscience, Psychology, Pharmacology and Children's Health), University of Florence, Italy
[2] Department of Educational Sciences and Psychology, University of Florence, Italy
[3] Centre for the Study of Complex Dynamics, University of Florence, Florence, Italy
[4] Department of Physics and Astronomy, University of Florence, Florence, Italy

**Abstract.** The main aim of our study was to analyze the effects of a virtual environment on social conformity, with particular attention to the effects of different types of task and psychological variables on social influence, on one side, and to the neural correlates related to conformity, measured by means of an Emotiv EPOC device on the other. For our purpose, we replicated the famous Asch's visual task and created two new tasks of increasing ambiguity, assessed through the calculation of the item's entropy. We also administered five scales in order to assess different psychological traits. From the experiment, conducted on 181 university students, emerged that conformity grows according to the ambiguity of the task, but normative influence is significantly weaker in virtual environments, if compared to face-to-face experiments. The analyzed psychological traits, however, result not to be relatable to conformity, and they only affect the subjects' response times. From the ERP (Event-related potentials) analysis, we detected N200 and P300 components comparing the plots of conformist and non-conformist subjects, alongside with the detection of their Late Positive Potential, Readiness Potential, and Error-Related Negativity, which appear consistently different for the two typologies.

## 1    Introduction

Conformity has been widely analyzed by social psychology starting from the pioneeristic works of Sherif and Asch [1].
These experiments showed to what extent majority pressure can be powerful, even when the majority is giving a clearly incorrect answer.
However, from an evolutionary point of view, these results are not shocking, since conformity turns out to be an adaptive behavior that presents many benefits concerning human beings' fitness, reproduction and survival [2].
Recent cultural studies on conformity, analyzed its connection with protection and showed how the inhabitants of areas that historically had higher prevalence of disease tend to be more conformist, and this outcome is explained by the fact that conformity is a strong protective factor against the risk of contracting illness [3].

Among the different benefits, conformity can work as a protective shield against threats linked to group exclusion: infact human beings developed heuristics and neurally evolved with the ability to select similar individuals to bond with and to distinguish in-group members from out-group members. From this point of view, conformist behaviors like mimicry can be helpful in creating group membership [4].

The context plays a crucial role in fostering this type of behavior but the majority of the studies on conformity focused on face-to-face interaction.

However, considering the widespread of social networks and computer-mediated communication nowadays, it is necessary to shed light on how social influence works in a context characterized by anonymity.

Contrasting theoretical frameworks focused on the effects of anonymity on human interaction: on one side anonymity seems to be able to give individuals a feeling of protection that leads them to feel more free to speak their minds [5], but on the other side, the Social Identity Model of Deindividuation Effects (SIDE) perspective shows how anonymity can lead to deindividuation, and this factor, making less salient individual traits, can lead to a stronger tendency to conform to social norms [6].

From the very few studies on conformity in virtual environments emerged that social influence can occur also in virtual environments but with some differences according to the type of influence elicited.

A replication of Asch's experiment, showed no conformity in anonymous condition [7]. Asch's task, which consisted in confronting a reference bar with three options of different lengths, among which was present only one twin bar, is an example of normative influence, namely the tendency to conform in order not to appear as an outsider when confronting a group [8]. Asch's experimental organization consisted in a group of seven people among which only the person in sixth answer position was the experimental subject. In some trials, the majority was asked to provide unanimously the same incorrect answer, and the tendency to conform of the experimental subject was analyzed. Averagely, 32% of the experimental subjects conformed to the majority [1]. In this case, since the task's ambiguity was low and detecting the correct answer was pretty easy, the reason that brought the subjects to conform is relatable to in-group dynamics, social norms and the desire not to break them [8].

The existing neurophysiological literature deepens the construct of normative influence, showing how Event-Related Potentials (ERP, measurable brain responses resulting from a specific cognitive, sensory or motor event) components such as N200 (that is a negativity associated with a variation in form or context of a predominant stimulus, typically evoked between 180 and 325 ms following the presentation of a particular visual or auditory stimulus) and P300 (which is a positivity typically emerging approximately between 300 and 400 ms following the stimulus presentation, and perhaps the most-studied ERP component in research concerning selective attention and information processing) respectively indicate the internal conflict experienced by the subject and the activation of inhibitory response mechanisms [9], as well as the awareness of the conformity of the response [10] [11].

Other experiments that used more ambiguous or difficult tasks, showed how informational influence, namely the tendency to conform when the subjects have lacking information on the task and for this reason reckon the group a reliable source, can occur also in virtual environments [12].

In this case, group dynamics are less relevant, since the goal is to give a correct answer [8].

The aim of the present study was to analyze how normative and informational influence could be affected by a virtual context, so the effect given by the type of task was taken into account. Besides replicating Asch's visual task, we created two more tasks whose items presented different levels of ambiguity, assessed with the measurement of the item's entropy.

The cultural items consisted in a target work associated with three adjectives with different levels of semantic relation with the target word, while the apperceptive items consisted in invented words associated with existing adjectives and vice-versa.

The experiment was also conducted in different conditions, the first one concerning different levels of anonymity, in order to see if a higher exposure could have an effect on conformity, and the second one making the subjects perform the experiment alone or with the physical presence of other subjects in the same room.

For our purpose, we created a virtual interface that simulated the responses of six non-existing people, with the experimental subject placed always in the sixth response position, in order to be able to see the responses of a majority of five subjects, inside a group of seven people.

Besides these variables, we also controlled the interaction with personality traits, in order to analyze whether it is possible to predict conformity from certain psychological features.

We performed the same experiment on subjects wearing an Emotiv EPOC device, a wireless EEG-based headset that enables the detection of electrical brain signals on the scalp's surface, in order to record and analyze the ERP components.

The ERP experiment focused on the differences between conformist and non-conformist subjects' cerebral activity within all the tasks.

Besides analyzing N200 and P300, ERP components such as LPP (Late Positive Potential), RP (Readiness Potential) and ERN (Error-Related Negativity) were reckoned to be potentially interesting for the phenomenon taken into consideration, because they respectively indicate emotional regulation [13], premotor planning of voluntary movement [14] and error awareness [15].

## 2    Participants

For our study we recruited 181 universitary students: 120 participated to the standard experiment and 60 participated to the ERP version of the study.

For the experimental typologies, we balanced them for the full and partial anonymity conditions and the group and single condition.

The only unbalanced condition is gender, with 139 females and 42 males, but in the data analysis phase, the factor has been controlled. The recruitment took place through voluntary census and the majority (80%) came from the School of Psychology of the University of Florence.

## 3      Method and procedure

To control the possible effect of psychological variables on conformity, the first experimental phase consisted in a preliminary survey composed of a battery of self-reported socio-demographic and psychological questionnaires and scales. After this phase, the subjects performed the experiment on a software that re-created a group condition. Finally they were asked to fill a questionnaire investigating their experience within the group.

**Materials.** The preliminary survey was composed by two sections. The first section consisted in socio-demographic items concerning age, gender, type of studies attended, educational level, marital status, presence of children and religious orientation. The second section consisted in a series of scales investigating psychological traits and status, in particular the scales administered were:

- The Fast Five Personality Questionnaire [16]
- The State-Trait Anxiety Inventory for Adults [17]
- The Multidimensional Sense of Community Scale [18]
- The Rosenberg's Self-Esteem Scale [19]
- The General Self-Efficacy Scale [20]

**Software.** To perform the experiment we created a software designed on Google Script, the functioning of which is similar to Crutchfield's [21] apparatus. Before starting, the experimental subjects were informed that six other subjects were about to log in and participate with them. After reading the instructions, they could log-in. The interface was organized in order to simulate the responses of six non-existing people, with randomized log-in and response times. The interface provided also the possibility to manipulate anonymity and to collect the subjects' response times.
On the left was placed a series of dots, vertically numbered from 1 to 7, associated with each group member: in the fully anonymous condition, the subjects could only see the numbers associated with the response order, while in the partially anonymous condition, they could see names and surnames. The experimental subjects were always placed in the sixth position, so that they could see the answers provided by five fake subjects before: when a subject answered, a number indicating their choice appeared beside their number or name, and when it was the experimental subject's turn, the stimulus appeared as long as three buttons (numbered 1, 2 and 3) in correspondence of the three alternatives. The first task was Asch's adaptation, with twenty items, the second the cultural and the third the apperceptive, each composed by forty-five items.
This second phase took averagely forty-five minutes to be completed.

**Setting.** The experiment was presented as a study on visual and semantic perception.
We collected the contacts and scheduled the appointments with the subjects via e-mail or text message, making sure that they fit the non-psychological disease condition.
The subjects were then randomly assigned to the experimental conditions.
The group-condition experiments took place in the computer science laboratory, with groups of six, seven or eight people.
They were equipped with headphones playing white noise and each workstation had a barrier isolating each subject.
The single-condition and ERP experiments took place in the social psychology laboratory, where the subjects performed the expriment alone with the presence of maximum three experimenters. Each experiment lasted approximately an hour and twenty minutes.

## Data analysis

The first step consisted in a pre-processing of the data, in which we verified the necessary pre-conditions for the inferential analysis.
For the *t*-Student analysis, we balanced the experimental conditions for each subgroup, made sure to have the necessary minimum numerosity, and verified a proper gaussian distribution for the dependent variables. Where necessary, we proceeded with a re-normalization of the dependent variables by means of a logarithmic function.
In order to calculate the entropy of each item in cultural and apperceptive tasks, we presented the items to two samples of people (71 subjects for the cultural and 79 for the apperceptive) without manipulation, collected their responses, and calculated the frequencies and percentages of the answers to each item.
On the basis of the percentages we calculated the entropy for each item "*i*" using the equation 1, with $p_j^k = (\sum_{i=1}^{n} r_i^k)/n$, "*n*" equal to the number of respondents, and "$r_i^k$" reporting the answer of the subject "*i*" to the item "*k*" (i.e., that can be "0" or "1").

$$E^k = \sum_{j=1}^{3} - p_j^k \log p_j^k \qquad (1)$$

When we collected all the entropies, we calculated the median for cultural and apperceptive tasks and according to that, we divided the items in high and low entropy. In order to assess the effect of entropy on the decision making we balanced the distribution of the entropy within the set of experimental stimuli.
We adopted a *t*-Student test for independent samples to analyze the relations between conformity, delay and type of anonymity.
To analyze the relations between type of task, conformity, delay and entropy, we proceeded with a $\chi^2$ test. To analyze the psychological and socio-demographic variables effects we conducted a *r* Pearson's correlations and again a $\chi^2$ analysis.
ERP data were properly filtered using Matlab in order to be analyzed and then *t*-Student tests were adopted.

## 4 Results

Starting from the anonymity effect, below are presented the significant values. The affected variables are the general delay and the delay in conformist answers in Asch's task, the general delay in the cultural task and conformity in the cultural task (Table 1).

**Table 1.** Anonymity effect on conformity and delay. *= $p < .05$

| Condition | Anonymity | Mean | St. Dev. | t |
|---|---|---|---|---|
| General delay (Asch) | No Anonymity | 12438.60 | 41050.13 | 1.98* |
|  | Anonymity | 3781.48 | 1207.95 |  |
| Conformity-related delay (Asch) | No Anonymity | 10576.88 | 30842.82 | 2.03* |
|  | Anonymity | 3911.59 | 1191.88 |  |
| General delay (Cultural) | No Anonymity | 6500.87 | 3441.42 | 2.04* |
|  | Anonymity | 5649.43 | 1909.53 |  |
| Conformity (Apperceptive) | No Anonymity | 0.27 | 0.14 | -1.99* |
|  | Anonymity | 0.33 | 0.21 |  |

For what concerns the type of task, significant relations appeared between all the investigated factors, which are conformity, entropy and delay (Table 2, Table 3)

**Table 2.** Type of task effect on conformity and delay. ***= $p < .001$; [1]=4805ms

|  |  | Task |  |  | $\chi^2$ |
|---|---|---|---|---|---|
|  |  | Asch | Cultural | Apperceptive |  |
| Conformity | No | 98.6% | 84.8% | 70.2% | 954.64*** |
|  | yes | 1.4% | 15.2% | 29.8% |  |
| Delay | < median[1] | 59.8% | 60.2% | 56.9% | 19.15*** |
|  | > median[1] | 40.2% | 39.8% | 43.1% |  |

**Table 3.** Entropy effect on conformity and delay. ***= $p < .001$; [1]=0.427; [2]=4805ms

|  |  | Entropy |  | $\chi^2$ |
|---|---|---|---|---|
|  |  | < median[1] | > median[1] |  |
| Conformity | No | 92.6% | 69.6% | 1065.396*** |
|  | Yes | 7.4% | 30.4% |  |
| Conformity delay | < median[2] | 60.5% | 55.9% | 25.272*** |
|  | > median[2] | 39.5% | 44.1% |  |

The correlations between conformity, delay and psychological traits did not provide significant results, only state and trait anxiety have an impact on the delays in the cultural task's responses. For what concerns the socio-demographic variables, no gender differences in conformist behavior appeared.

The ERP plots obtained by using Emotiv EPOC were divided according to two specific times of the test, namely considering two trigger-moments. In the patterns relating to the stimulus administration presented below, emerged how typical ERP components such as N200, P300 and LPP (Late Positive Potential) are significantly higher for the conformist subjects (Figure 1).

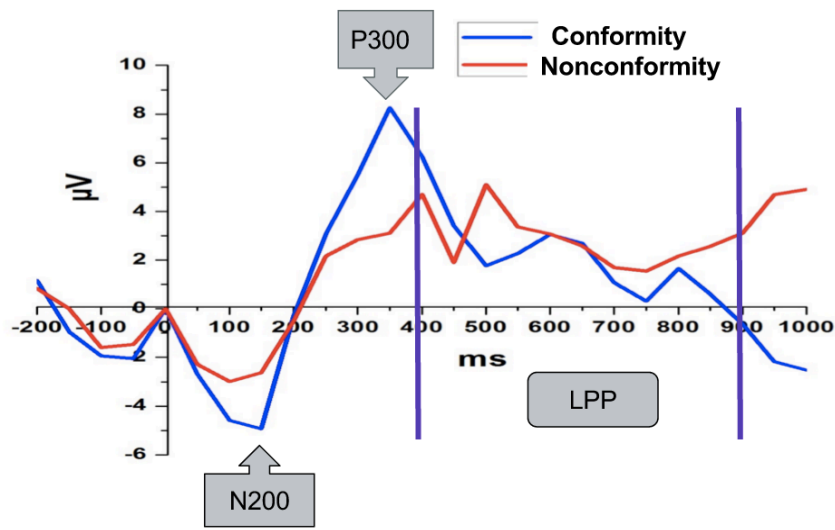

**Fig. 1.** Grand-averages of conformist and non-conformist plots elicited by stimulus administration. Left hemisphere electrodes (no occipital): E1, E2, E3, E4, E5. Y-axis reports the cerebral activation in microvolts, while x-axis reports the time in milliseconds from the stimulus administration. Blue line stands for conformist subjects, red line for non-conformist subjects. N200 indicates a negativity typically evoked 180 to 325 ms following the presentation of a stimulus (i.e. mismatch detection), P300 is a positivity that peaks approximately 300 - 400 ms post-stimulus (i.e. selective attention), finally LPP begins around 400 ms after the onset of a stimulus and lasts for a few hundred milliseconds (i.e. emotional activity).

**Table 4.** Observable components' means, standard deviations and t-Students of conformist and non-conformist plots relative to the time of stimulus administration. *** = $p < .001$

| Observable | Condition | Mean | St. Dev. | t |
|---|---|---|---|---|
| N200 | Conformist | -4,06 | 5,18 | 38,80*** |
| | Non-conformist | -2,10 | 5,17 | |
| P300 | Conformist | 5,29 | 5,17 | 19,74*** |
| | Non-conformist | 2,94 | 5,18 | |
| LPP | Conformist | 0,64 | 5,18 | -15,95*** |
| | Non-conformist | 2,97 | 5,19 | |

Conversely, in the plots related to the click time, which represented the final decision made by the subject, it clearly emerges how components such as RP (Readiness Potential) and ERN (Error-Related Negativity) are significantly higher for the non-conformist subjects (Figure 2).

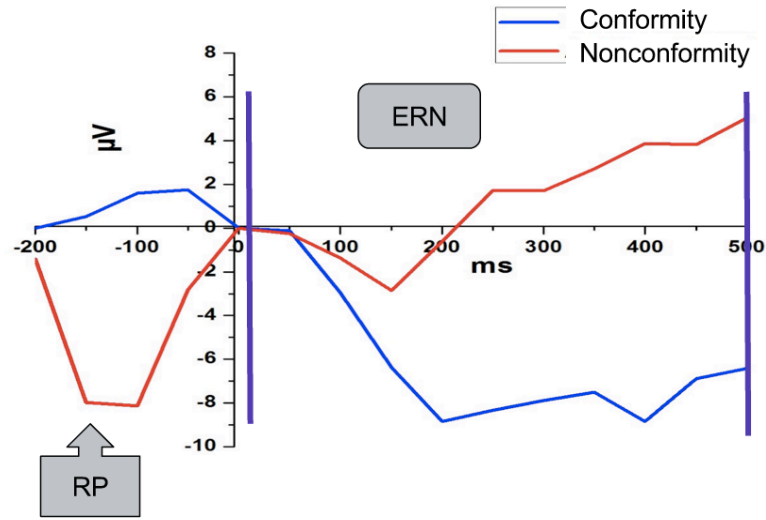

**Fig. 2.** Grand-averages of conformist and non-conformist plots following the click's time. Frontal electrodes: E1, E3, E12, E14. RP peaks around 250 – 0 ms before the time of movement execution (i.e. voluntary movement planning), and ERN is a negativity beginning around the first 50 – 100 ms after the click (i.e. error awareness).

**Table 5.** Observable components' means, standard deviations and t Students of conformist and non-conformist plots relative to the time of the click, *** = $p < .001$.

| Observable | Condition | Mean | St. Dev. | t |
|---|---|---|---|---|
| RP | Conformist | 0,97 | 5,19 | 31,54*** |
|  | Non-conformist | -5,08 | 5,52 |  |
| ERN | Conformist | -6,42 | 5,23 | -36,14*** |
|  | Non-conformist | 1,38 | 5,53 |  |

## 5  Discussion

This research, aimed at highlighting the differences that may be elicited by a virtual environment in social phenomena, specifically conformity and its interaction with other factors.

At first, we obtained different percentages of conformity according to the type of task: in Asch's task, only 1.4% of the subjects conformed when the majority gave an unanimous incorrect answer, while the percentages grow in the cultural (15.2%) and in the apperceptive (29.8%). These preliminary results suggest that Asch's paradigm changes in a virtual, anonymous environment, and that normative influence might be less effective due to the characteristics of the setting and social norms might take longer to become effective.

The percentages emerged in the cultural and apperceptive tasks, however, showed a growth of conformity with more ambiguous items.

Furthermore, the $\chi^2$ analysis showed a strong connection between conformity and entropy, suggesting that the higher is the entropy, the higher is the tendency to conform, inducing informational influence.

Another interesting factor consisted in the relationship between delay, conformity and type of task provided by the $\chi^2$ analysis. The results showed how the delay in the responses tended to be generally longer when the subjects conformed to the majority's opinion, and also that the delay increased in the cultural and apperceptive tasks, so when the ambiguity of the stimulus was higher.

For what concerns anonymity, the only significant result on the effect of the type of anonymity on conformity appeared in the apperceptive task, while the other results showed an effect on the different types of delay in relationship with the task. The single and group conditions presented no significant differences.

For what concerns the psychological and socio-demographic factors, the correlations appeared to be weak.

Finally, we confirmed the existing ERP literature investigating N200 and P300, adding other potentially interesting components (LPP, RP, ERN): N200 indicated the incongruence between subjects' answer and group answer, P300 the behavior adjustment, LPP the consequential emotional regulation, and all of them resulted more pronounced in the conformist subjects. RP probably indicates the premotor click planning, ERN the awareness of the error committed, and these components are more evident in the case of the non-conformist subjects. Thanks to such identifications, we have been able to differentiate the conformist events from the nonconformist ones in the electroencephalographic patterns. The ERP analysis, for now, considered only the differences between conformists and non-conformists, further analysis will focus on the ERP components related to entropy and type of task.

**Acknowledgements.** This work was supported by EU Commission (FP7-ICT-2013-10) Proposal No. 611299 SciCafe 2.0.

## References

1. Asch, S. E. (1951). Effects of group pressure upon the modification and distortion of judgments. *Groups, leadership, and men. S*, 450-501.
2. Morgan, T. J. H., & Laland, K. N. (2012). The Biological Bases of Conformity. *Frontiers in Neuroscience*, *6*, 87. doi:10.3389/fnins.2012.00087

3. Murray, D. R., Trudeau, R., & Schaller, M. (2011). On the origins of cultural differences in conformity: Four tests of the pathogen prevalence hypothesis. *Personality and Social Psychology Bulletin*, *37*(3), 318-329.
4. Neuberg, S. L., Kenrick, D. T., & Schaller, M. (2010). Evolutionary social psychology. *Handbook of social psychology*.
5. Kiesler, S., Siegel, J., & McGuire, T. W. (1984). Social psychological aspects of computer-mediated communication. *American psychologist*, *39*(10), 1123.
6. Spears, R., Postmes, T., Lea, M., & Wolbert, A. (2002). When are net effects gross products? Communication. *Journal of Social Issues*, *58*(1), 91-107.
7. Laporte, L., van Nimwegen, C., & Uyttendaele, A. J. (2010, October). Do people say what they think: Social conformity behavior in varying degrees of online social presence. In *Proceedings of the 6th Nordic Conference on Human-Computer Interaction: Extending Boundaries* (pp. 305-314). ACM.
8. Deutsch, M., & Gerard, H. B. (1955). A study of normative and informational social influences upon individual judgment. *The journal of abnormal and social psychology*, *51*(3), 629.
9. Chen, J., Wu, Y., Tong, G., Guan, X., & Zhou, X. (2012). ERP correlates of social conformity in a line judgment task. BMC neuroscience, 13(1), 43.
10. Kim, B. R., Liss, A., Rao, M., Singer, Z., & Compton, R. J. (2012). Social deviance activates the brain's error-monitoring system. Cognitive, Affective, & Behavioral Neuroscience, 12(1), 65-73.
11. Shestakova, A., Rieskamp, J., Tugin, S., Ossadtchi, A., Krutitskaya, J., & Klucharev, V. (2013). Electrophysiological precursors of social conformity. Social cognitive and affective neuroscience, 8(7), 756-763.
12. Rosander, M., & Eriksson, O. (2012). Conformity on the Internet–The role of task difficulty and gender differences. *Computers in human behavior*, *28*(5), 1587-1595. (12)
13. Dennis, T. A., & Hajcak, G. (2009). The late positive potential: a neurophysiological marker for emotion regulation in children. *Journal of Child Psychology and Psychiatry*, *50*(11), 1373-1383.
14. Libet, B., Gleason, C. A., Wright, E. W., & Pearl, D. K. (1983). Time of conscious intention to act in relation to onset of cerebral activity (readiness-potential). *Brain*, *106*(3), 623-642.
15. Klein, T. A., Endrass, T., Kathmann, N., Neumann, J., von Cramon, D. Y., & Ullsperger, M. (2007). Neural correlates of error awareness. *Neuroimage*, *34*(4), 1774-1781.
16. Giannini, M., Pannocchia, L., Lauro-Grotto, R., Gori, A. (n.d.) A measure for counseling: the five factor adjective short test (5-fast). Unpublished manuscript, Department of Psychology, University of Florence, Florence, Italy
17. Spielberger, C. D., & Gorsuch, R. L. (1983). State-Trait Anxiety Inventory for Adults: Sampler Set: Manual, Tekst Booklet and Scoring Key. Consulting Psychologists Press.
18. Prezza, M., Pacilli, M. G., Barbaranelli, C., & Zampatti, E. (2009). The MTSOCS: A multidimensional sense of community scale for local communities. *Journal of Community Psychology*, *37*(3), 305-326.
19. Rosenberg, M. (1979) Conceiving the self. In Ciarrochi, J., & Bilich, L. (2007). Acceptance and Commitment Therapy. Measures Package. *Unpublished manuscript, University of Wollongong, Wollongong, Australia*
20. Sibilia, L., Schwarzer, R., and Jerusalem, M.: Italian adaptation of the general self-efficacy scale. Resource document. Ralf Schwarzer web site (1995)
21. Crutchfield, R. S. (1955). Conformity and character. *American Psychologist*, *10*(5), 191.